# Spontaneous Chaotic Microlasers for Random Number Generation

Chun-Guang Ma, Jin-Long Xiao, Zhi-Xiong Xiao, Yue-De Yang, and Yong-Zhen Huang*

State Key Laboratory of Integrated Optoelectronics, Institute of Semiconductors, Chinese Academy of Sciences, Beijing 100083, China & Center of Material Science and Optoelectronic Technology, University of Chinese Academy of Sciences, Chinese Academy of Sciences, Beijing, China

*Corresponding author E-mail: yzhuang@semi.ac.cn

**Abstract**

Chaotic semiconductor lasers have been widely investigated for generating true random numbers with the merit of unpredictability relative to pseudorandom numbers generated by a computing process, especially for the lasers with external optical feedback. Here, we report the first spontaneous chaotic microlaser due to the interaction of nearly degenerate modes. The circular-side hexagonal microlaser is designed and demonstrated with adjusted mode intervals in the order of GHz for the fundamental and the first order transverse modes. Four modes including degenerate modes with strong mode beating will result in chaos. True random numbers at 10Gb/s is obtained from the laser output power with a high dimension of 5.46. Our finding provides a novel and easy way to create controllable and robust optical chaos for high speed random number generation, which also paves the way for the nonlinear dynamics in a solitary laser.

The random numbers are crucial in the generation of cryptographic keys for classical and quantum cryptography systems, the reliability of modern networked society and stochastic simulation, etc[1-4]. Pseudo-random numbers can be generated using deterministic algorithm programs, but with limited unpredictability, which reduces the security and randomness. Physical entropy sources, such as electronic and photon noises, thermal noise in resistors, frequency jitter in oscillators, have been applied to generate non-deterministic truly random number sequences[5-7]. However, the generation rates of the non-deterministic random number sequences from the stochastic processes are less than 100 Mbit/s[8], due to the low signal levels of physical noises.

High-bandwidth chaotic semiconductor lasers have been widely investigated for the generation of truly random numbers[8-25] and secure communications[26-29]. Semiconductor lasers are extremely sensitive to external perturbations because their lasing frequency is affected by carrier density due to a non-zero linewidth enhancement factor. Chaotic semiconductor lasers can be realized using external optical feedback[8-14] and optical injection[23, 24, 30, 31], and integrated lasers with a passive feedback cavity or a mutual coupling integrated laser[15-21, 32]. However, time-delay periodicity imprinted in the laser output leads to recurrences in the outcome for the chaotic lasers with external optical feedback[33, 34]. And post-processing is usually required to obtain truly random numbers. Random number sequences at a rate of 1.7 Gbps with verified randomness were generated by combing the fluctuating optical outputs of two independent chaotic lasers[9]. The high derivative of the digitized output intensity of a chaotic laser with external feedback was used to generate random sequences at rates up to 300 Gbit/s[8]. In addition, low-dimension chaos in polarization-resolved output power was reported for a free-running quantum-dot vertical-cavity surface-emitting laser[35, 36], which was caused by chaotic polarization mode hopping due to nonlinear mode competition including carrier spin relaxation. Chaotic solitary laser is the most prominent configuration for the random number generation due to its robust and simple scheme.

In this article, we demonstrate a spontaneous chaos microlaser due to the interaction among nearly degenerate multiple modes for a circular-side hexagonal microresonator (CSHM) semiconductor laser. Similar as microdisk resonators, polygonal resonators can also support high Q whispering-gallery modes with the mode light rays totally internal reflection at side walls of the polygon, such as square and hexagon resonators[37, 38]. A dual-mode lasing square microlaser was designed and fabricated with a ring p-electrode, which can introduce a refractive index step and tune transverse-mode interval[39]. Furthermore, we designed circular-side square and hexagon resonators to enhance mode $Q$ factor and adjust transverse mode interval[40-42]. Dual-transverse-mode lasing square resonator microlasers with mode intervals up to sub-THz were realized, due to ultrahigh passive mode $Q$ factors and different period lengths for different transverse modes caused by the circular sides. The fundamental and first order transverse modes with high passive $Q$ factors can form dual-mode lasing with near threshold gain determined by the other losses, such as material absorption loss. Here, we design a circular-sides hexagon microresonator to enhance mode $Q$ factors and adjust mode frequency interval. The transverse mode interval can be reduced to the order of GHz, which are corresponding to different longitudinal modes. Accounting degenerate modes, we can have four modes with mode intervals in the order of GHz, which can result in strong mode interaction inside the microresonator and induce chaotic output. Because the carrier density can follow the variation of intensity beating between modes in the order of GHz. A chaotic deformed hexagon microlaser is demonstrated experimentally, and truly random numbers at 10 Gb/s obtained from the output intensity of the microlaser are verified by statistical tests.

**Mode simulation for deformed hexagonal microresonators**

A deformed hexagon as shown in Fig. 1a in *x-y* plane was simulated for TE modes by two-dimension finite-element method (FEM), where $R$ is a radius of circular side, $a$ is the flat side length of original hexagon, $w$ is the width of an output waveguide connected to a vertex at an angle of $\theta$, and $\delta$ is a deformation parameter. The effective refractive index was taken to be 3.2 for an AlGaInAs/InP multiple quantum wells laser resonator, which is confined by a polymer layer with a refractive index of 1.54.

For a deformed hexagon with $a = 10$ μm, $w = 1.5$ μm and $\theta = 55°$, the *z*-direction magnetic field (|Hz|) distributions are plotted in Figs. 1b and 1c for the fundamental and first order transverse modes $H_0$ and $H_1$ at wavelengths of 1551.6 and 1551.5 nm, respectively, of degenerate modes with higher Q-factor at $\delta = 1.015$ μm. The mode field patterns are confined well with very weak distribution at vertices of the hexagon and small radiation loss, due to the concave mirror effect of the circular sides. Mode wavelengths versus deformation parameter $\delta$ are calculated and plotted in Fig. 1d, where $H_0$ and $H_1$ have near the same wavelength, but with a longitudinal mode number difference of 2. The corresponding mode $Q$ factors are presented in Fig. 1e at $\delta = 1.015$ μm, for all the high-$Q$ modes, where two degenerate modes of the fundamental transverse modes $H_0$ have $Q$ factors around $6.0 \times 10^5$ and two of the first order transverse modes $H_1$ are near $1.0 \times 10^4$. Furthermore, a ring-patterned electrode is simulated to finally tune the mode wavelength interval by assuming a refractive index step $\Delta n$ in the ring region as shown in the inset of Fig 1a. The transverse mode wavelength interval versus $\Delta n$ is plotted in Fig. 1f at the ring with a width of 5 μm and an external radius of 9.5 μm. The results indicate that the transverse mode intervals can be adjusted by the refraction index step as the square microresonator[39].

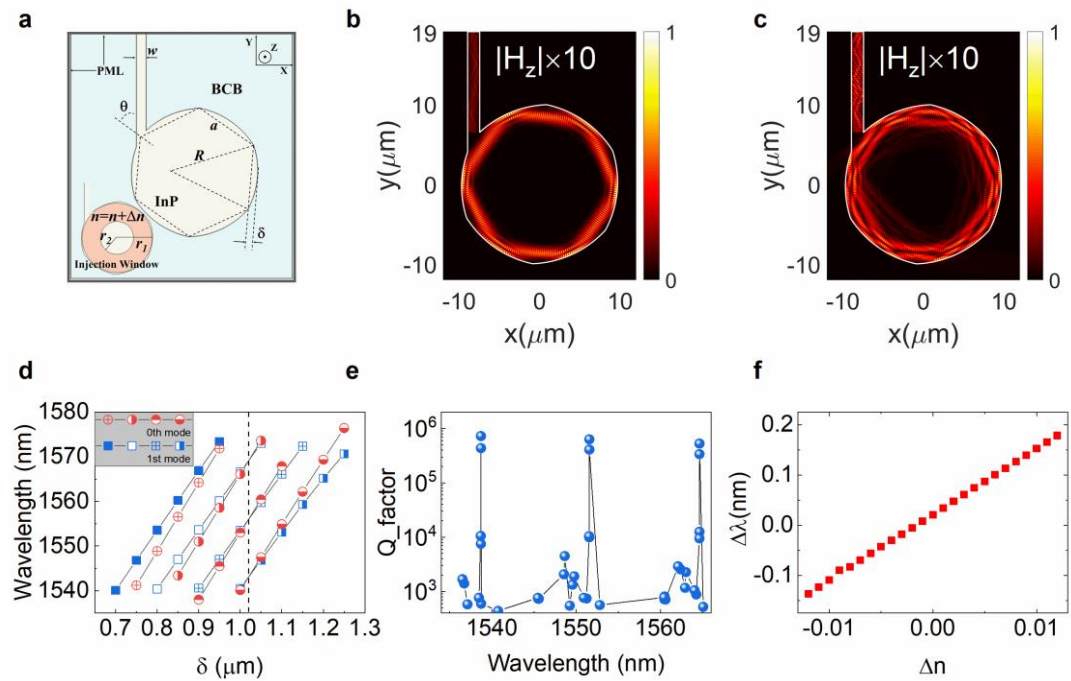

**Figure 1 Simulated mode characteristics for a deformed hexagon with** $a = 10$ μm, $w = 1.5$ μm and $\theta = 55°$. **a,** Schematic diagram of the simulated hexagon resonator with a tilted output waveguide. Inset: the resonator with a ring-patterned injection window. **b, c,** Mode field distributions of |Hz| for the fundamental and first order transverse modes at 1551.6 and 1551.5 nm, respectively. **d,** Resonant wavelength versus $\delta$ for the fundamental transverse mode $H_0$ (circle symbols) and the first order transverse mode $H_1$ (square symbols). Solid, open, cross and half-filled

symbols are corresponding to the same longitudinal mode number, respectively. **e**, Mode $Q$ factors versus mode wavelengths at $\delta = 1.015$ μm for high $Q$ modes including degenerate modes. **f**, The mode wavelength interval between the fundamental and the first order transverse modes versus the refractive index step $\Delta n$ in the ring electrode region.

Accounting degenerate modes, we have four high $Q$ modes marked as $H_{0,1}$ and $H_{0,2}$, and $H_{1,1}$ and $H_{1,2}$, respectively, with the first and second subscripts represent the transverse mode number and degenerate mode number. To model practical devices, we simulate deformed hexagon resonator with sidewalls modulated by a random fluctuation. Taking the fluctuation magnitude of 50 nm, we have mode $Q$ factors about $0.3 \times 10^5$ and $0.8 \times 10^4$ for $H_0$ and $H_1$ modes, respectively, and the mode frequency interval in the order of GHz for split degenerate modes. In addition, vertical radiation loss[43] will reduce mode Q factor difference again. For a microcavity laser under an optical injection, light intensity inside the microcavity laser was modulated at a detuning frequency between the injected light and lasing mode, and the corresponding carrier density oscillating at the detuning frequency was verified by measuring microwaves from the laser electrode[44]. The obtained microwave power versus the microwave frequency was agreement very well with small-signal modulation curve of the free-running microsquare laser, and different dynamic states, such as period-one and period-two oscillation and chaos were verified from lasing spectra and microwave spectra. In addition, four-wave mixing peaks were observed in a dual-mode lasing microlaser with a mode interval tuning from 30 to 50 GHz [39], which reveals mode interaction inside the microlaser without external optical injection. So plentiful nonlinear dynamic behaviors are expected for a microlaser with multiple modes with mode interval in the order of GHz, as the carriers can oscillate at mode frequency intervals. In supplement, we simulate a semiconductor laser with four modes and observe near chaotic output by a rate equation model with a beating intensity term [45]

**Lasing characteristics of chaotic microcavity laser**

A deformed hexagon microlaser as shown in Fig. 2a with $a = 10$μm, $d = 1.5$μm, $\delta = 1.015$μm and $\theta = 55°$ was tested at a temperature of 287K controlled using a thermoelectric cooler. The multi-mode-fiber (MMF) and single-mode-fiber (SMF) coupled powers as well as the applied voltage versus continuous-wave injection current are plotted in Fig. 2b, with a threshold current about 6mA and the maximum powers of 39 and 62 μW coupled into SMF and MMF, respectively. The output power spectra, which was measured by a high-speed photonic detector (PD) and an electrical spectrum analyzer, are plotted in Fig. 2c at the injection current $I = 22$, 23, and 24 mA with the noise spectrum measured at $I = 0$. Three evident peaks with heights less than 10 dB are observed at the frequencies of 1, 2.5, and 11.5 GHz at 22 mA.

Detailed lasing spectra around 1552 nm from 19 to 26 mA are plotted in Fig. 2d, which shows evident broadening spectra with multiple wide peaks. From 22 to 24 mA, four peaks with intensity difference less than 2dB are observed in a near flat band with mode intervals ranged from 0.03 to 0.07 nm (frequency intervals 4 to 9 GHz), which should be the fundamental and first order transverse modes including degenerate modes. Multiple-mode lasing with such small intervals can result in complicated nonlinear process due to carrier oscillation and mode competition. As shown in Fig. 2e, another group of modes around 1566 nm become main lasing modes with the increase of injection current due to the heating effect. At 21, 24 and 25 mA, the lasing spectra are similar as those in Fig. 2d with wide chaotic peaks determined by the main lasing modes around 1552 nm. But at 27, 29, 33, and 37, two main modes are observed with mode intervals of 0.16, 0.18, 0.19, 0.18 and 0.13 nm, respectively, with evident four-wave mixing peaks. At 44 mA, additional peaks of period-two oscillate state appear due to the small mode frequency interval of 16 GHz between the two main modes, as carriers can oscillate with a large amplitude following mode beating intensity at 16 GH. For the period-one and period-two oscillate states, microwave spectra have narrow peaks with a height up to 30 dB, respectively[46]. The whole lasing spectra are plotted in Fig. 2f to show the lasing mode jumping with the increase of injection current.

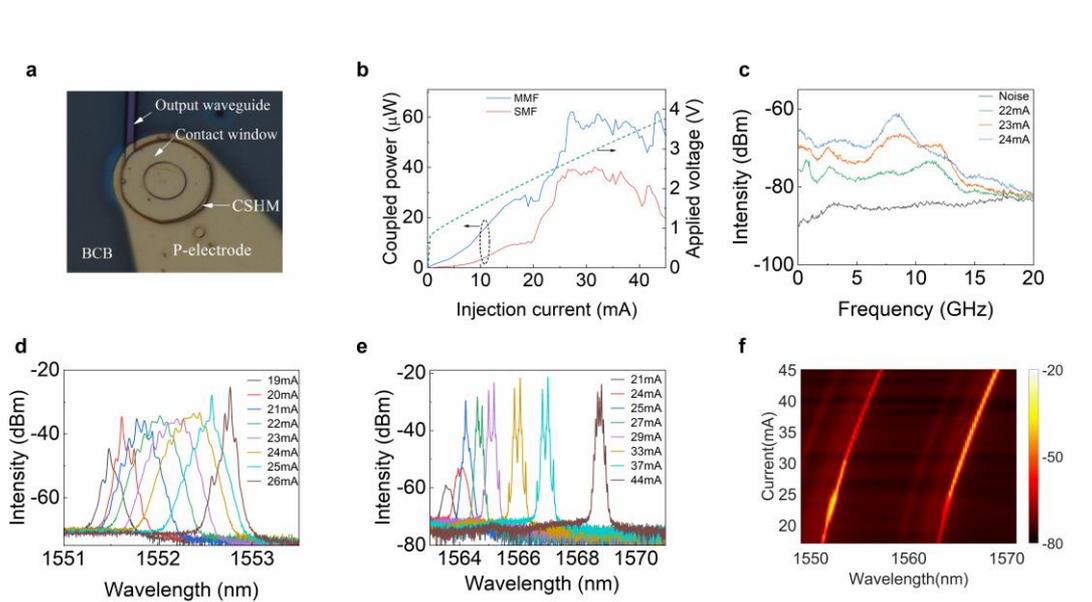

**Figure 2 Lasing characteristics of a deformed hexagon microlaser.** a, Microscopic image of a fabricated microlaser. **b**, Output power coupled into a SMF (red) and MMF (blue), and applied voltage versus injection current. **c**, Measured microwave spectra from laser output at 22, 23 and 24 mA. **d**, **e** Detailed lasing spectra around 1552 nm and 1566 nm at different currents. **f,** Lasing spectra in a range covering two longitudinal modes changing with the variation of I from 17 to 45 mA.

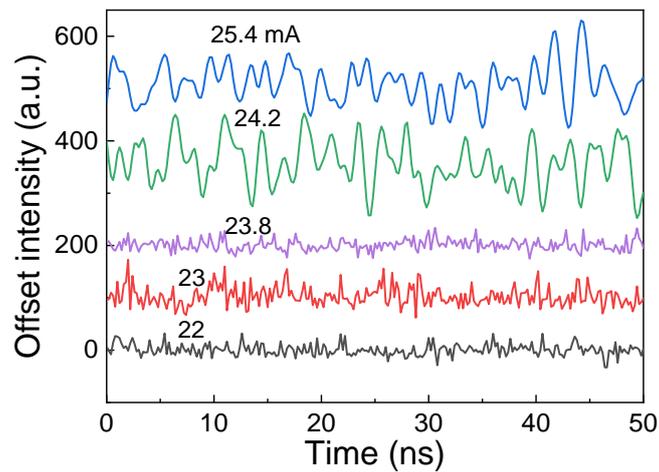

**Figure 3 Chaotic output of the deformed hexagon microlaser.** Temporal a.c. waveform of the laser output signals at 22, 23, 23.8, 24.2 and 25.4mA obtained at 5GSa/s.

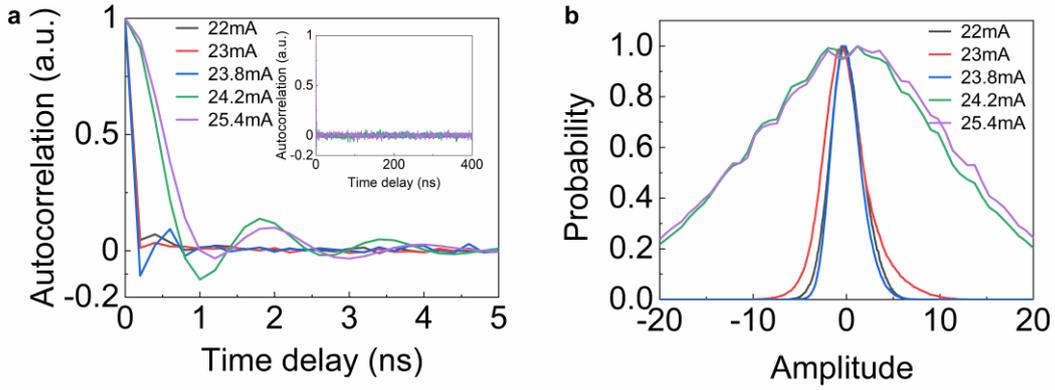

**Figure 4 Statistical behavior of the chaotic output of the deformed hexagon microlaser.**
**a,b**, Autocorrelation functions and probability functions calculated from 10000 samples at 22, 23, 23.8, 24.2, and 25.4 mA, respectively. The inset in (a) shows autocorrelation functions over 100 ns.

## Chaos output and identification

The output of the deformed hexagon laser was amplified by an erbium doped fiber amplifier and converted to an electrical signal by a 30-GHz-bandwidth photodetector with a.c. coupling to receive only the a.c. component. The a.c. signal was then measured by an oscillator with 2.5 GHz-bandwidth and 8-bit ADC. Temporal a.c. signal waveforms at 22, 23, 23.8, 24.2 and 25.4 mA were plotted in Fig. 3, which were measured by the oscillator at a sampling rate of 5GSa/s. The temporal waveforms show noise-like intensity oscillation in sub-nanosecond scale at 22 to 23.8 mA and nanosecond scale at 24.2 and 25.4 mA, respectively, which are corresponding to the wide and narrow chaotic lasing spectra as comparing Fig. 2d and Fig. 3. Autocorrelation functions and intensity probability distributions of the temporal waveforms are plotted in Figs. 4a and 4b, which are calculated from the output waveforms of 10000 samples. The autocorrelation function has a centre peak with a full width at half maximum (FWHM) of 1 ns and decayed oscillation peaks extended to about 4 ns, at 24.2 and 25.4 mA. But the autocorrelation functions at 22, 23 and 23.8 mA are near Dirac delta functions with a FWHM less 0.2 ns and evident minor peaks below 1 ns. The intensity probability distributions are Gauss distributions at 22, 23 and 23.8 mA with a much narrower width and smoother distribution than that at 24.2 and 25.4 mA.

Finally, the chaotic output data are evaluated using the modified version of the Grassberger & Procaccia's algorithm to estimate the correlation dimension $D_2$ and the $K_2$-entropy (Kolmogorov entropy) [47,48]. For the time series data with $N$ points, we resemble them into $d$-dimensional $N-d$ vectors spanned by a time delay step $\tau$ with $d$ as the embedding dimension[49]. Then correlation integral $C_d(r)$ of the new sequence vectors with the embedding dimension $d$ is calculated as the average number of vectors inside a ball of a radius $r$ around a single vector, where the radius is calculated as the distance between the two $d$-dimensional vectors. Because the numbers are different for different vectors, the average number is calculated by averaging the numbers around all the vectors. For chaotic data, $C_d(r)$ should converge and give the estimated correlation dimension $D_2 = \ln C_d(r)/\ln(r)$ and the $K_2$-entropy $K_2 = \ln\left(C_d(r)/C_{d+1}(r)\right)/\tau$ as $d$ and $r$ approach infinity and zero, respectively[48]. The chaotic output data can be assigned as periodic or quasiperiodic, chaos, and purely stochastic as $K_2$ is zero, positive, and infinity, respectively[47,48]. The calculated correlation integral versus the embedding dimension are presented in

Fig. 5a from the chaotic output data at 25.4 mA with $N$=40000. The correlation integral $C_d(r)$ converges with the increase of $d$, and the correlation dimension $D_2$ converges to about 5.46 as shown in Fig. 5b, by fitting the slope of the curve [50]. Furthermore, the $K_2$-entropy of $K_2 \approx 3.3$ ns$^{-1}$ is obtained as shown in Fig. 5c, which verifies the chaotic characteristics of the laser output data.

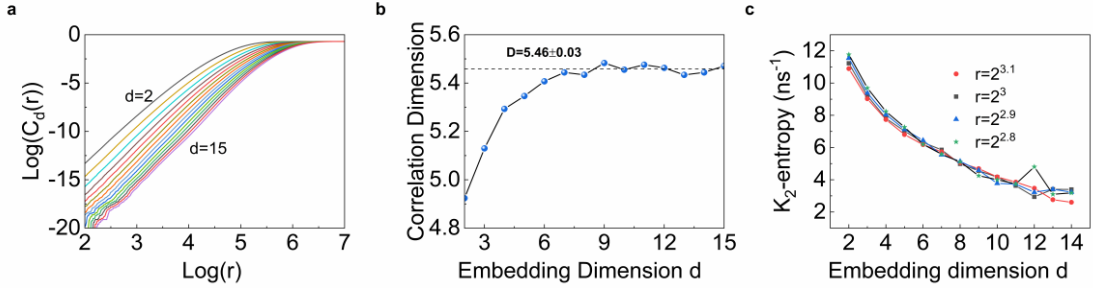

**Figure 5 Characteristics of the chaotic output for the deformed hexagon microlaser. a,** Correlation integral $C_d(r)$ versus sphere radius $r$. **b,** The estimated correlation dimension $D_2 = \ln C_d(r)/\ln(r)$ versus the embedding dimension $d$. **c,** The calculated $K_2$-entropy versus the embedding dimension $d$ obtained at the different radius $r$.

From the temporal waveforms, random bits were generated by retaining 2 the least significant bits (LSBs) of each sample for the microlaser at 25.4 mA. The generated random bit sequences are verified by the NIST Special Publication 800-22 statistical tests. A total of 120 sequences with each size of 1 Mbit were collected and tested. As summarized in Table 1, random bits generated at a bit-rate of 10 Gbit/s (5GS/s×2bits) are passed the test.

Table 1. Results of the NIST Special Publication 800-22 statistical tests.
The random bits of 120 sequences with each size of 1 Mbit were tested, and the worst P-value and proportion were presented.
At significance level α=0.01, the success proportion should be in the range of 0.99 ±0.027 and the composite P-value should be larger than 0.0001 to ensure uniformity.

| Statistical Test | P-value | Proportion | Result |
|---|---|---|---|
| Frequency | 0.1481 | 0.975 | Success |
| Block-frequency | 0.6371 | 1.000 | Success |
| Cumulative-sums | 0.8343 | 0.967 | Success |
| Runs | 0.2041 | 0.975 | Success |
| Longest-run | 0.9320 | 1.000 | Success |
| Rank | 0.3925 | 0.992 | Success |
| FFT | 0.9705 | 0.983 | Success |
| Nonperiodic-templates | 0.2536 | 0.967 | Success |
| Overlapping-templates | 0.9643 | 1.000 | Success |
| Universal | 0.3781 | 1.000 | Success |
| Approximate-entropy | 0.3115 | 0.983 | Success |
| Random-excursions | 0.2509 | 0.975 | Success |
| Random-excursions-variant | 0.0078 | 0.975 | Success |
| Serial | 0.6718 | 0.983 | Success |
| Linear-complexity | 0.4220 | 0.983 | Success |

**Discussion**

In summary, we report the first chaotic microlaser caused by internal mode interaction and the generation of real random numbers from the microlaser output with a chaos dimension of 5.46. By introducing circular sides and ring electrode for deformed hexagon microresonators, we can have four high-Q modes with mode frequency intervals in the order of GHz, i.e., the fundamental and first order transverse modes and their degenerate modes. The mode beating terms between different modes can result in locally intensity oscillation at the frequencies of mode intervals and cause the oscillation of carrier density by stimulated emission. Plentiful nonlinear dynamic states of chaos, period-two and period-one states were observed for the deformed hexagon microlasers with the increase of the mode intervals adjusted by injection current. High-dimension chaotic state with true random numbers at 10Gb/s was realized using the laser output of the deformed hexagon microlaser. Comparing the correlation functions in Fig. 4a, we expect that true random numbers can be realized at 50~100 Gb/s at 22 to 23.8 mA, i.e., 5~10 times of that at 25.4 mA. The deformed hexagon microcavity lasers provide a more simple, robust and solitary chaotic laser source than chaotic lasers caused by external optical injection or optical feedback.

Enhancing mode Q factors and tuning mode intervals by using circular sides and ring electrode pave the way for mode engineering in polygonal microcavities. Four high Q modes in the deformed hexagon microlaser can all approach threshold condition if their passive mode Q factors are much larger than that determined by absorption loss of laser wafer. The oscillations of four modes will result in local distributions of mode beating intensities inside the microcavity, and carrier density will follow the oscillation of the beating intensity locally as the mode frequency interval in the order of GHz. Furthermore, we can expect mode interaction similar as mode injection effects between different modes. Mode interaction processes will be very complex for a microcavity with four high-Q modes at such small mode intervals. Accounting mode beating terms, we present photon number as the sum of four modes and smaller beating terms between modes in the rate equation of carrier density as a whole ignoring local variation inside the microcavity. In such a simple rate equation model with four high-Q modes, we can get near chaos laser output waveform as shown in Supplement. However, the simulated output waveform shows stronger correlation than the experimental results in Fig. 4 with the correlation function decaying much slowly. In fact, mode beating terms oscillate with time and position inside the microcavity, which requires a space dependence rate equation model, in addition the time-variation.

**Methods**

**Simulation details**

The mode characteristics of the deformed hexagon micro resonator were simulated by two-dimensional finite element method (FEM) (COMSOL Multiphysics 5.0) with a cell step of 25 nm. The hexagon resonator is laterally confined by divinylsiloxane-bisbenzocyclobutene (DVS-BCB) with a refractive index of 1.54. A perfectly matched layer (PML) is used to terminate the calculation region, and the distances between the PML and the resonator edge are larger than 2μm to ensure the accuracy of the numerical calculation. TE modes is simulated which is the domination polarization for the compressively stressed laser wafer.

**Experimental data acquirement**

The microlasers were fabricated using an AlGaInAs/InP compressively strained multiple-quantum-well laser wafer as in [37]. The lasers were cleaved over the output waveguide and bonded p-side up on an AlN sub-mount. Lasing spectra were measured by coupling into a SMF and using an optical analyzer (OSA) with a resolution of 0.02 nm. The output of the microlaser was coupled into a tapered SMF and amplified using an erbium-doped fiber amplifier, then the amplified output optical signal was fed into a 30 GHz high-speed photodetector for the microwave signal generation. The microwave signal was measured using a 26.5 GHz bandwidth electric spectrum analyzer.

**Supplement**

**Rate equation simulation with four-mode beating terms**

To verify the nonlinear dynamics influenced by mode beating, a comprehensive rate equation model including four-mode beating terms was utilizing to analyze output behaviors:

$$\frac{dn}{dt} = \frac{\eta I}{qV_a} - An - Bn^2 - Cn^3 - v_g g(n, S_t) S_t, \tag{S1}$$

$$\frac{ds_m}{dt} = [\Gamma v_g g(n, S_t) - \alpha_i v_g] s_m - \frac{s_m}{\tau_{mpc}} + \Gamma \beta Bn^2, \tag{S2}$$

$$S_t = \sum_{m=1}^{4} s_m + k \sum_{i=1}^{4} \sum_{j=1}^{4} \sqrt{s_i s_j} \cos(2\pi \delta f_{ij} t) \quad i \neq j, \tag{S3}$$

where $n$ and $s_m$ are the carrier density and the photon density of the *mth* modes, $S_t$ is the average total photon density including beating terms, $\delta f_{ij} = f_i - f_j$ is the mode frequency difference. $I$ is the injection current, $\eta$ is current injection efficiency, $q$ is the electron charge, $V_a$ is the volume of the active region for the microlaser, $v_g = c/n_g$ is the group velocity of the lasing mode, $\beta$ is the spontaneous emission rate, and $A$, $B$, and $C$ are the defect, bimolecular, and Auger recombination coefficients, respectively. $\alpha_i$ is the internal material absorption loss and $\tau_{mpc} = Q_m/\omega_m$ is the mode lifetime determined by the passive mode quality factor $Q_m$. $\Gamma$ is the optical confinement factor which gives the spatial overlap between the active region volume and the mode volume. The coupling coefficient $k$ is set to indicate the partial overlap of the mode intensity between different modes.

The parameters used in the simulation were taken the same as Ref. 50. The rate equations (S1)-(S3) were solved by using a fourth-order Runge-Kutta integration with a time step of 200 fs in time domain to generate the time series output, and Fourier transforms were used to produce the optical and RF signal spectra., The passive mode Q factors were taken to be 7200, 7300, 7000 and 7100 for modes 1 to 4, respectively. The mode field was calculated from the mode photon density[49], and total mode field was expressed as the sum of the mode fields of the four modes ignoring mode phase differences except that caused by mode frequency differences. The simulation results were shown in Fig. S1 at $\delta f_{2,1}$ = 8 GHz, $\delta f_{3,1}$ = 3 GHz, $\delta f_{4,1}$ = 10.9 GHz, $k$ = 0.3 and $I$ = 15mA. Calculated optical spectra in Fig. S1a show the four main modes and additional peaks caused by the nonlinear process of mode beating, and the RF spectra in Fig. S1b have multiple peaks with main peaks around 2.9 and 8.1 GHz. The time domain series of output optical power as squared total mode field were shown in Fig. S1c, which shows aperiodic oscillation in sub-nanosecond scale. However, the autocorrelation coefficient of the output optical power in Fig. S1d shows weak periodic trend with a much longer correlation time than that shown in Fig. 4a. In fact, mode beatings are local effect inside microcavity based on mode field distributions. A simplified beating term is accounted in the rate equation model of (S1)-(S3). A space-dependent rate equation model is required to account the practical mode beating effect.

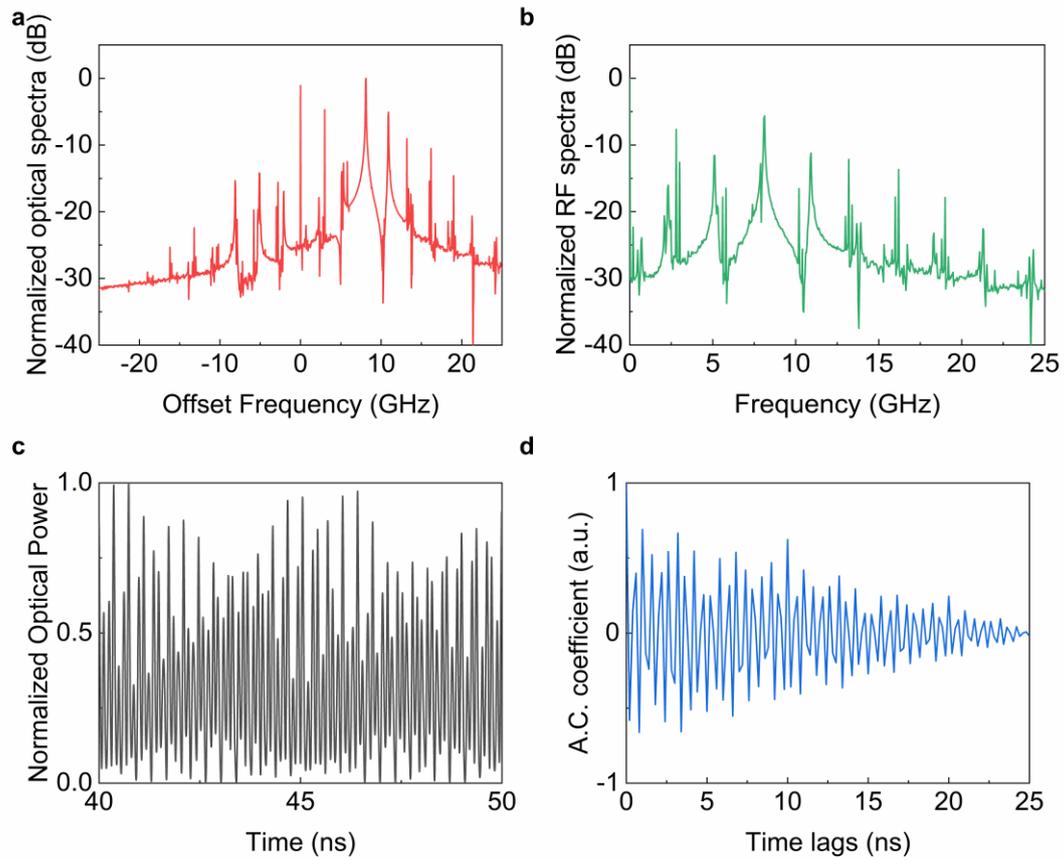

Figure S1 **Simulation output results for four modes** at $\delta f_{2,1}$ = 8 GHz, $\delta f_{3,1}$ = 3 GHz, $\delta f_{4,1}$ = 10.9 GHz *k*=0.3 and *I*=15mA. **a, b,** Simulated output optical spectra and RF spectra. **c**, Temporal series of output optical power. **d**, Autocorrelation coefficient of the output optical power.